\documentclass[10pt]{article}
\usepackage[letterpaper]{geometry}
\usepackage{hicss}
\usepackage{times}
\usepackage[none]{hyphenat}
\usepackage[hyphens]{url}
\usepackage{latexsym}
\usepackage{indentfirst}
\usepackage{graphicx}
\graphicspath{{images/}}
\usepackage{natbib}

\title{Surveying the Operational Cybersecurity and Supply Chain Threat Landscape when Developing and Deploying AI Systems}

\author{Michael R. Smith \\
 Sandia National Labs\\
 {\underline{ msmith4@sandia.gov}} \\ \And
 Joe Ingram\\
 Sandia National Labs \\
 {\underline{ jbingra@sandia.gov} } \\ }

\date{}

\begin{document}
\maketitle
\begin{abstract}
The rise of AI has transformed the software and hardware landscape, enabling powerful capabilities through specialized infrastructures, large-scale data storage, and advanced hardware.
However, these innovations introduce unique attack surfaces and objectives which traditional cybersecurity assessments often overlook. 
Cyber attackers are shifting their objectives from conventional goals like privilege escalation and network pivoting to manipulating AI outputs to achieve desired system effects, such as slowing system performance, flooding outputs with false positives, or degrading model accuracy. 
This paper serves to raise awareness of the novel cyber threats that are introduced when incorporating AI into a software system.
We explore the operational cybersecurity and supply chain risks across the AI lifecycle, emphasizing the need for tailored security frameworks to address evolving threats in the AI-driven landscape.
We highlight previous exploitations and provide insights from working in this area. 
By understanding these risks, organizations can better protect AI systems and ensure their reliability and resilience.
\end{abstract}

\subsubsection*{Keywords:}

Artificial Intelligence, Cybersecurity, Adversarial Machine Learning, Supply Chain

\section{Introduction}

The ascension of artificial intelligence (AI) has profoundly transformed the software and hardware landscape, driving advancements across industries and enabling innovative solutions to complex problems. 
AI systems rely on specialized software infrastructures and run-time environments to optimize intricate mathematical operations as well as large-scale data storage to manage AI models and data both on disk and in memory.
Likewise, advancements in specialized hardware, such as graphical processing units (GPUs) and tensor processing units (TPUs), have enabled the efficient execution of computationally-intensive tasks. 
These advancements have collectively contributed to the remarkable effectiveness and efficiency of AI systems.

However, the same components that make AI systems powerful also introduce unique and novel attack surfaces that can be exploited by malicious actors.
Further, the AI portion is only one component in a larger system.
From vulnerabilities in data pipelines to risks in hardware and software supply chains, the operational and cybersecurity concerns surrounding AI systems are multifaceted and demand specialized attention.
Despite the growing prevalence of AI, traditional cybersecurity assessments often fail to adequately address the specific risks associated with AI systems and the associated changes in objectives by attackers. 
This oversight leaves critical gaps in security, exposing AI systems to threats that can compromise their integrity, reliability, and safety.
This paper surveys many of the operational cybersecurity and supply chain concerns inherent in the development and deployment of AI systems. 

We distinguish this work from the large and growing body of literature on adversarial AI that exploits algorithmic vulnerabilities to manipulate the behavior of AI models.
These attacks target weaknesses in the data, training process or inference mechanisms.
For example, evasion attacks perturb inputs causing a model to make incorrect predictions~\citep{szegedy2013intriguing}.
We provide a brief overview and how it relates to cyber vulnerabilities---specifically the motivation of an attacker.
Further, adversarial AI as presented in many academic venues is often removed from operational considerations and may fail in real-world situations~\citep{zhang2022operational, grosse2024towards}.

Cyber attacks on the AI portion of a system are often more practical to achieve than the mathematical adversarial AI attacks (assuming cyber access) \citep{apruzzese2023real}.
We examine vulnerabilities across the AI threat landscape including the learned models, specialized data, software infrastructure, run-time environment, and hardware, highlighting the unique challenges and research in each.
By identifying these risks and proposing mitigation strategies, this paper aims to provide a comprehensive overview for assessing the security of AI systems against evolving threats in an increasingly AI-driven world.

The following section provides an overview of what we mean by an AI system.
Section~\ref{sec:CAML} briefly overviews adversarial AI, the objectives that can be adopted by cyber attackers, and assumptions in executing attacks.
Section~\ref{sec:conops} connects the objectives of adversarial AI and how it changes the concept of operations (CONOPS) of a cyber attack on an AI system.
We then survey the AI attack landscape and how these objectives can be met by an attacker in each component of the AI threat landscape.
We share examples in the academic literature, proofs-of-concept presented in cyber blogs and reports, and experiential insights where applicable.
Section~\ref{sec:lifecycle} presents the various stages of the AI development-to-deployment lifecycle and where these attacks can occur.
We conclude with a discussion of the future of security in AI systems.

\section{Brief Overview of AI Systems, Large Language Models, and Agentic Systems}
\label{sec:llms}

\begin{figure*}[thb]
    \centering
	\includegraphics[clip,width=0.9\linewidth]{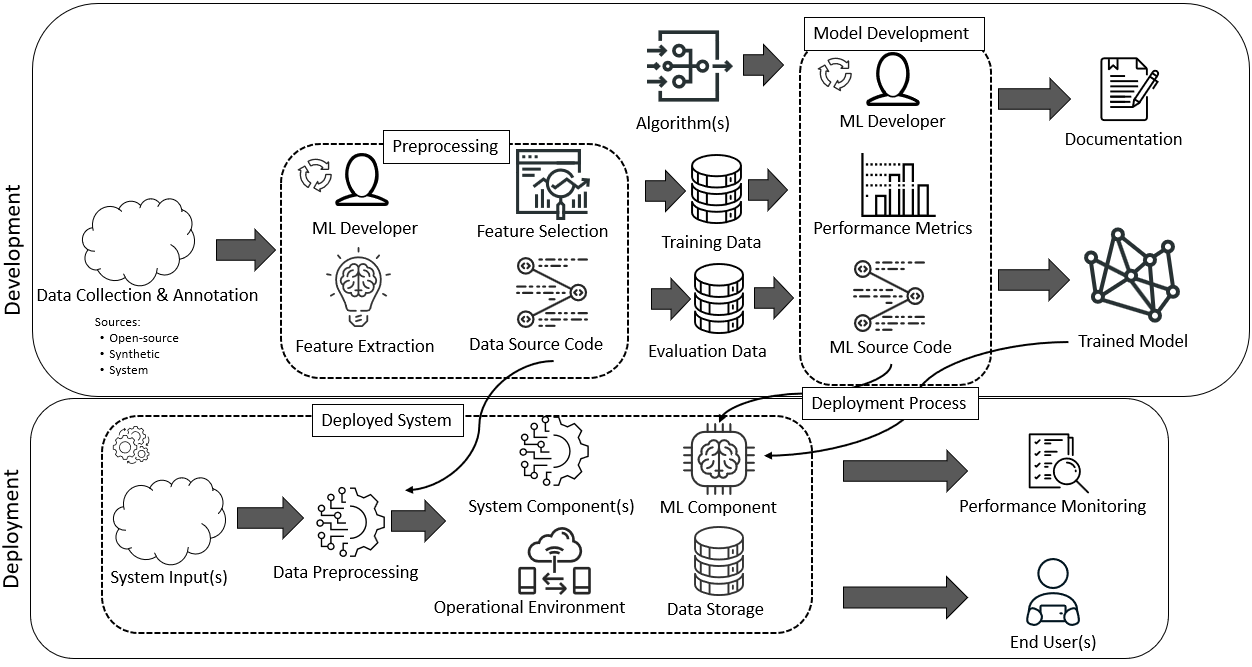}
	\caption{Typical AI lifecycle comprised of data curation and labeling, preprocessing, model development, deployment and operation of the AI system. Each phase exposes an additional attack surface.}
	\label{fig:lifecycle}       
\end{figure*}

When we discuss AI systems, we refer to software and hardware systems where the AI represents a single component of a larger system.
Figure~\ref{fig:lifecycle} shows an exemplar AI lifecycle from development to deployment. 
While considerable effort is exerted to develop the model in the development stage, when deployed, the AI model is a software component connected within a larger system of sensors, data storage, and other system components.
The AI component can be anything that is data-driven including complex object detection models or relatively simpler linear/logistic regression models.

In natural language processing, AI has evolved significantly.
Large Language Models (LLMs), such as OpenAI's GPT~\citep{brown2020language} or Google's Gemini~\citep{team2024gemini}, are advanced AI models trained on large corpora of text data to understand and generate human-like language. 
These models excel in tasks such as text summarization, translation, question answering, and code generation.

Agentic systems extend the capabilities of LLMs by integrating autonomous decision-making and execution mechanisms. 
These systems are designed to perform complex workflows with little-to-no human intervention.
They leverage LLMs for reasoning and communication while interfacing with external tools, APIs, and environments. 
For instance, an agentic system tasked with cybersecurity monitoring might autonomously analyze network logs, detect anomalies, and initiate mitigation actions. 
While autonomy enhances efficiency and scalability, it also introduces cybersecurity risks. 
For example, attackers can exploit vulnerabilities in input validation, API dependencies, or decision logic to manipulate the system’s actions or outcomes.

Together, LLMs and agentic systems exemplify the growing sophistication of AI technologies, offering transformative capabilities while presenting new challenges in security. 
As these systems become increasingly integrated into critical applications, addressing their vulnerabilities will be paramount for ensuring their safe and effective deployment.

\section{Brief Background on Adversarial AI and Attack Assumptions}
\label{sec:CAML}
There is a growing body of adversarial attacks on AI to exploit vulnerabilities in models to manipulate their behavior, degrade performance, leak private information, or achieve other malicious objectives. 
These attacks typically have a narrow focus on the model or data and target weaknesses in data, training processes, or inference mechanisms.
They pose a significant risk to AI systems deployed in security-critical applications.
However, most are based on mathematical computations of gradients---which may not be true attack vectors in practical attacks on AI systems~\citep{apruzzese2023real}.
These same objectives can be achieved by cyber attackers.

\begin{figure}[thb]
    \centering
	\includegraphics[clip,width=0.9\linewidth]{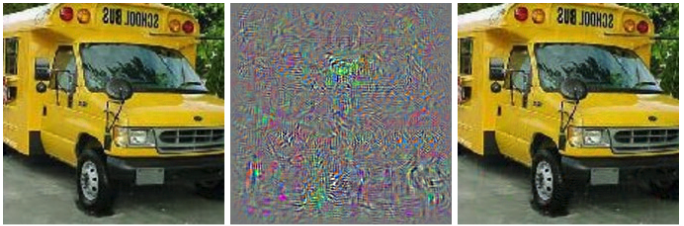}
	\caption{Evasion attack example. Left: Clean image that is correctly predicted as a school bus. Middle: Adversarial noise. Right: Attacked images which is a combination of the left and middle images and is predicted to be an "ostrich". From~\citet{szegedy2013intriguing}.}
	\label{fig:evasion}       
\end{figure}

Evasion attacks craft adversarial inputs that cause incorrect predictions during inference. 
These attacks are typically executed by introducing subtle perturbations to input data, imperceptible to humans but sufficient to manipulate the model’s decision-making as shown in Figure~\ref{fig:evasion}. 
The techniques used in evasion attacks depend on the attacker’s level of access to the model. 
In white-box attacks, the attacker has full knowledge of the model’s architecture, parameters, and gradients, enabling precise manipulation of inputs. 
For example, the Fast Gradient Sign Method (FGSM)~\citep{goodfellow2014explaining} calculates the gradient of the loss function with respect to the input and applies perturbations in the direction that maximizes the loss. 
In contrast, black-box attacks rely on querying the model to infer its behavior without direct access to internal details. 
Techniques like the Hop-Skip-Jump Attack~\citep{chen2020hopskipjumpattack} iteratively refine adversarial inputs by estimating decision boundaries using minimal queries.

\begin{figure}[thb]
    \centering
    \begin{tabular}{c}
	\includegraphics[clip,width=0.9\linewidth]{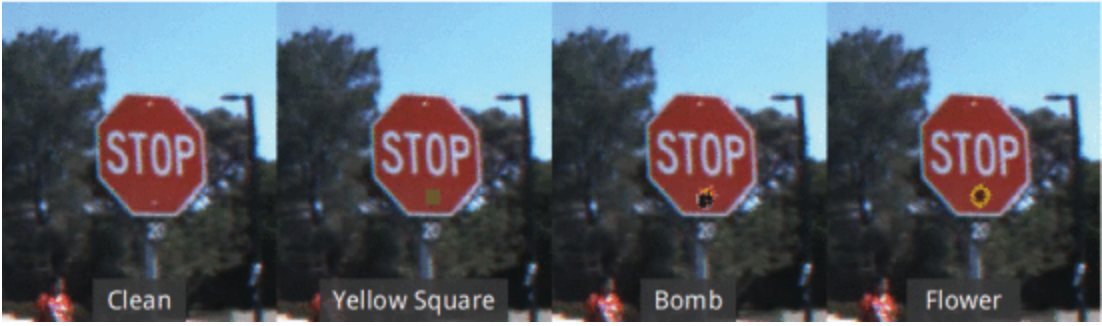}\\
 (a) \\
	\includegraphics[clip,width=0.5\linewidth]{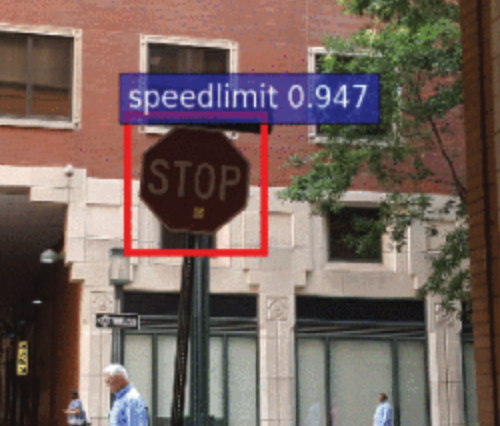}\\
 (b)
    \end{tabular}
	\caption{Data poisoning attack. (a) Examples of a clean image (far left) with three poisoned examples. (b) With a yellow square, the model predicts the stop sign to be a speed limit sign. From~\citet{gu2017badnets}.}
	\label{fig:poison}       
\end{figure}

Data poisoning targets the training phase of machine learning models by injecting malicious data into the training dataset to manipulate the model’s behavior---to degrade performance or embed vulnerabilities. 
One common technique is label flipping, where attackers modify the labels of specific data points (e.g., flipping the label of a benign sample to ``malicious'') to distort the model’s decision boundaries~\citep{biggio2012poisoning, kegelmeyer2015counter}.
Another prominent method injects poisoned samples with hidden triggers into the training data to create backdoors in the model. 
These backdoors allow the model to behave normally under most conditions but produce specific outputs when triggered by adversarial inputs~\citep{gu2017badnets}---as shown in Figure~\ref{fig:poison}.
Recently, it was shown that triggers can be inserted through label flipping only~\citep{jha2023label}.

\begin{figure}[thb]
    \centering
    \begin{tabular}{c}
	\scalebox{-1}[1]{\includegraphics[clip,width=0.6\linewidth]{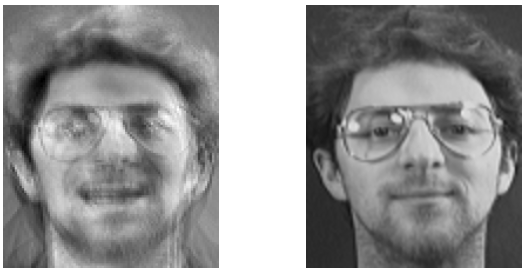}}\\
 (a) Original Image\ \ \ Extracted Image\\
	\includegraphics[clip,width=0.6\linewidth]{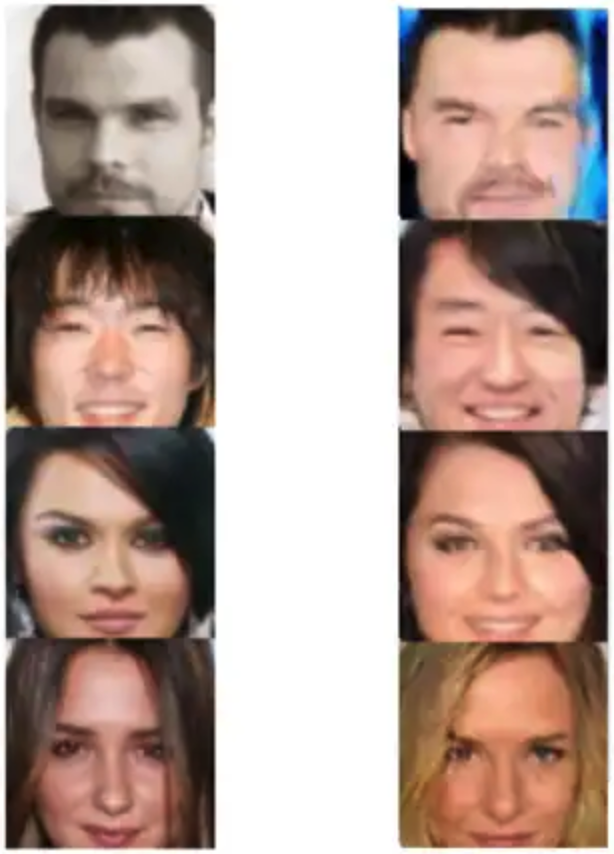}\\
 (b) Original Images\ \ Extracted Images \\
    \end{tabular}
	\caption{Model inversion attacks for stealing training data. (a) Original attack by~\citet{fredrikson2015model}. (b) More recent attacking showing significant progress in privacy attacks on AI~\citep{zhang2020secret}.}
	\label{fig:inversion}       
\end{figure}

Membership inference and model extraction attacks compromise privacy and intellectual property.
Membership inference attacks aim to determine whether specific data points were part of the model’s training dataset. 
Attackers typically query the model with data samples and analyze its confidence scores or output probabilities; higher confidence often indicates that a sample was part of the training set~\citep{shokri2017membership}.
Model extraction attacks, on the other hand, seek to replicate the functionality or parameters of a proprietary model or extract training data.
Techniques such as decision boundary estimation or gradient-based reconstruction allow attackers to infer the model’s architecture or weights~\citep{tramer2016stealing}.
A determined adversary can steal sensitive training data from an AI model, as demonstrated by~\citet{fredrikson2015model}.
An attacker can use model inversion techniques to reconstruct recognizable images of individuals from a facial recognition model.
See Figure~\ref{fig:inversion}.

LLMs provide additional attack scenarios.
Jailbreaking refers to manipulating LLMs to bypass built-in ethical, operational, or security safeguards. 
Attackers achieve this by crafting adversarial prompts or exploiting weaknesses in the model’s instruction-following mechanisms to elicit restricted behaviors.
For example, adversarial prompts can trick an LLM into providing instructions for malicious activities or circumventing content moderation filters~\citep{ganguli2022red}. 
As LLMs are integrated into critical applications, addressing jailbreaking risks is essential for ensuring their safe and ethical use---especially in agentic systems.

\section{CONOPS and Objectives for Cyber Attackers}
\label{sec:conops}
The traditional concepts of operations (CONOPS) and objectives for cyber attackers have historically focused on actions such as gaining unauthorized access, escalating privileges, pivoting within networks, and exfiltrating sensitive data.
The rise of AI systems introduces additional opportunities and motivations for attackers, fundamentally altering their objectives. 
Instead of solely targeting infrastructure, attackers can now focus on manipulating the behavior and outputs of AI systems as demonstrated in the previous section on adversarial AI. 
However, cyber access also allows for a broader disruption to the AI component.
For example, adversaries may seek to degrade the performance of an AI system by introducing delays or computational bottlenecks, effectively slowing down its ability to process and respond to inputs. 
Alternatively, attackers may aim to flood the system with false positives or adversarial inputs, overwhelming its decision-making processes and rendering its outputs unreliable.
These objectives can disrupt critical operations, erode trust in AI-driven systems, and even cause cascading failures in dependent systems. 
By targeting the unique vulnerabilities of AI systems—such as their reliance on data integrity, model accuracy, and computational efficiency—attackers are shifting their strategies to exploit the increasingly central role of AI in modern technology ecosystems.

\begin{figure}[thb]
    \centering
	\includegraphics[clip,width=0.9\linewidth]{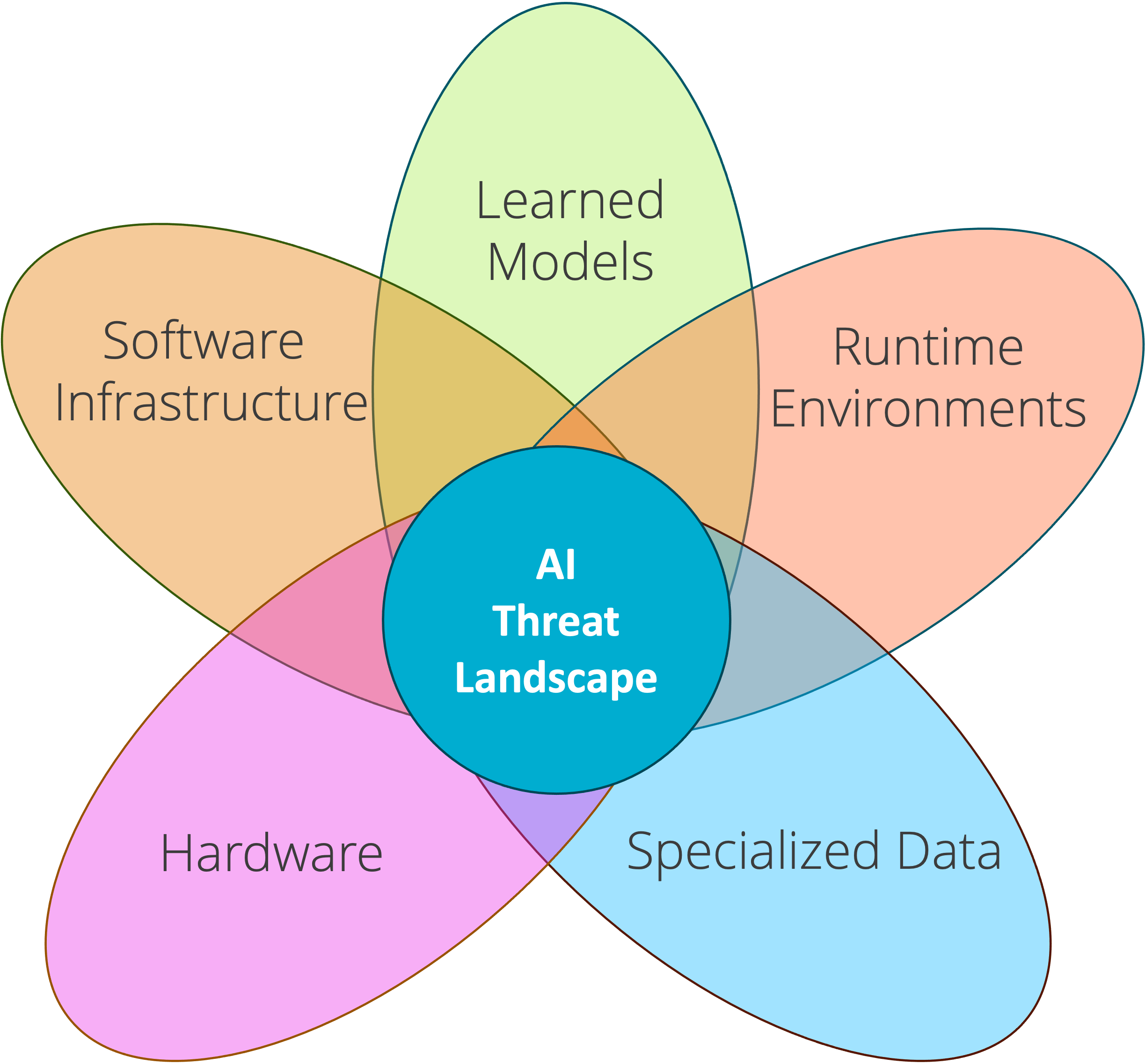}
	\caption{Venn diagram outlining the various components that enable AI and form the threat landscape.}
	\label{fig:landscape}       
\end{figure}

\section{AI Threat Landscape}

The same advancements in software, middleware, and hardware that enabled the success and efficiency of AI also expose additional attack surfaces.
We categorize the AI threat landscape encompassing (1)~learned models, (2)~specialized data, (3)~software infrastructure, (4)~run-time environments, and (5)~hardware as shown in Figure~\ref{fig:landscape}.

\subsection{Learned Models}
Pre-trained models, widely used in AI applications for their efficiency and ability to generalize across tasks, introduce significant cybersecurity risks. 
These models, often sourced from public repositories or third-party platforms, can serve as vectors for malicious activities, including poisoning attacks and malware introduction.
Attackers exploit the trust developers place in widely-used repositories. 

A poisoned model is one that has been intentionally manipulated during training to embed vulnerabilities or triggers as described in Section ~\ref{sec:CAML}.
Additionally, cyber attackers on an AI system could tamper with the model to alter performance~\citep{zehavi2023facial, grimes2024concept}.
Attackers can modify model parameters to achieve their desired objectives.

Pre-trained models can also contain malware.
Attackers may embed malicious code or scripts within the model files or associated dependencies.
For instance, downloading a compromised model from a platform like Hugging Face or PyTorch Hub could lead to the execution of malware.
In early 2024, more than 100 malicious models were discovered on Hugging Face~\citep{montalbano2024hugging}.
The malware exploited a deserialization vulnerability in the Pickle file that allows arbitrary code execution~\citep{milanov2024exploiting}.
Several mechanisms have been implemented to prevent the exploitation of this vulnerability.
However, as of early 2025, malicious models are \emph{still} found on Hugging Face exploiting this same vulnerability~\citep{zanki2025malicious}.
Part of the challenge is that despite knowledge of this vulnerability, alternative model sharing formats are limited and the AI community is focusing mostly on getting models to perform well rather than their security.
Until the security of a model is prioritized, it will continue to be a vulnerability.

Another vector for malware through AI models is through Lambda layers---lightweight, user-defined layers that allow developers to apply custom operations or transformations to input tensors using simple Python functions.
Lambda layers provide flexibility by leveraging the `marshal' Python module to serialize the lambda layer.
As is the same with Pickle files, the marshal module allows code to be executed during deserialization~\citep{lumelsky2024TensorFlowKeras}.
In the Keras library leveraged by Tensorflow, this vulnerability is addressed after version Keras 2.12 providing a ``safe mode."
However, to provide backward comparability, a downgrade attack is possible as Keras will load legacy, vulnerable code if the model file is from an older version of Keras ($<=2.12$).
Thus, even up-to-date AI libraries can still be exploited.

To address these risks, organizations should follow some simple guidance. 
Use cryptographic checksums or digital signatures to ensure the authenticity of pre-trained models.
Rely on reputable platforms and conduct audits of third-party contributions.
Analyze pre-trained models for embedded backdoors or malicious code using specialized tools.
Test models in isolated environments before integrating them into production systems.
Many AI libraries can enforce safe loads i.e., ensure that they are enabled.

\subsection{Specialized Data}
The foundation of any AI system is high-quality data. 
However, collecting and labeling a sufficient amount data is challenging and expensive for most organizations, and, hence, they rely on publicly or commercially available datasets.
This makes it particularly vulnerable to supply chain attacks or insider threats.
Combine this with the fact that many systems operate on complex data modalities, such as natural language, makes this stage highly vulnerable to attackers.

The reliance of LLMs and agentic systems on natural language as both input and output introduces significant vulnerabilities as natural language can serve as an attack vector.
Attackers can exploit LLMs to generate harmful code that is then posted back onto the system and executed with seemingly benign queries.
For example, LLM web applications can be vulnerable to cross-site scripting (XSS) attacks due to insecure output handling when the output from an LLM is posted back to the website that hosts the LLM~\citep{mik0w2023hackstery}.
Safeguards can be put into place; however, attackers can manipulate LLMs to bypass restrictions or safeguards to jailbreak the LLM.

If an agentic system is provided with too much autonomy, then malicious queries can result in cyber breaches.
SQL commands can be executed if connected to a database.
Code can be executed if the system has access to a terminal.
For example, a vulnerability in Microsoft 365's Copilot based on documents used to help guide the LLM (RAG or Retrieval-Augmented Generation) could exfiltrate data~\citep{french2025microsoft}.
Further, a vulnerability was disclosed in LangChain.
LangChain is a framework for integrating LLMs with external data sources, APIs, and workflows to create dynamic, context-aware AI systems.
The vulnerability (CVE-2023-29374) had a flaw in the `LLMMathChain' that enabled arbitrary code execution via the Python `exec' method~\citep{nist2023CVE202329374}.

The complexity and flexibility of natural language make it challenging to predict and defend against such attacks, underscoring the need for robust input validation, monitoring, and adversarial training to secure LLMs and agentic systems in cybersecurity-critical applications.
All automation should be limited in agentic systems.

\subsection{Software Infrastructure}
AI software frameworks, such as TensorFlow, PyTorch, and Keras, provide the foundation for building and deploying AI models. 
However, these frameworks can have vulnerabilities in their code, dependencies, and libraries, which can be exploited by attackers.

Attackers can compromise AI frameworks, libraries, or pre-trained models by embedding malicious code.
In late 2022, attackers leveraged a tactic known as ``dependency confusion" or ``repository hijacking" where an attacker impersonates a legitimate Python package or repository~\citep{constantinescu2023bitdefender}.
In this case, a malicious package named `torchtriton' was uploaded to the Python Package Index (PyPI), sharing the same name as a legitimate dependency in the PyTorch nightly package index.
The package manager, `pip', by default, prioritizes packages listed on PyPI over other indices when the extra-index-url argument is used. 
This allowed the malicious package to be installed instead of the legitimate version.
The compromised `torchtriton' contained a malicious binary called `triton' that collected and exfiltrated sensitive data, including usernames, IP addresses, environment variables, and specific files like ``/etc/hosts", ``/etc/passwd", and contents of the user's home directory.

Several additional vulnerabilities have been discovered in TensorFlow.
A series of continuous integration and continuous delivery/deployment misconfigurations in TensorFlow's GitHub pipeline would allow an attacker to compromise build agents to create malicious releases, remote code execution, and exfilatration of GitHub Personal Access Token~\citep{khan2024TensorFlow}.
While TensorFlow has addressed many concerns, certain vulnerabilities persist via a downgrade attack when TensorFlow will provide (vulnerable) functionality to older models to support legacy systems~\citep{lumelsky2024TensorFlowKeras}.
\citet{filus2023software} detailed several other vulnerabilities in TensorFlow and the difficulty of detecting them.

\subsection{Run-time Environment}
AI models require specialized run-time environments to interface with and leverage high-performance hardware, such as GPUs, TPUs, and embedded devices.
Training AI models, especially deep learning models, involves processing large datasets and performing billions of mathematical operations---necessitating high-performance computing environments.
However, these environments often differ from those where a model will be used.
For example, in autonomous vehicles, the hardware will differ from what it was trained on.

AI models deployed on edge devices, such as NVIDIA Jetson or Raspberry Pi, require lightweight environments tailored for low-power usage. 
Frameworks like TensorFlow Lite and ONNX Runtime enable efficient inference on embedded devices.
To enable deployment on commodity hardware, models trained on high-performance systems are often quantized or compressed. 
Quantization reduces the precision of model parameters (e.g., from 32-bit floating-point to 8-bit integers), while compression techniques like pruning remove redundant parameters. 
These processes significantly reduce model size and computational requirements but may slightly impact accuracy.

The process of converting a full-precision model to another version exposes an attack surface.
The academic literature has explored how the quantization can be exploited for LLMs~\citep{egashira2024exploiting} and object detection models~\citep{pan2021understanding, hong2021qu, tian2022stealthy}.
In these cases, the artifacts from compressing and quantizing the AI models serve as the triggers to manipulate the behavior of the AI models in nefarious ways.

\subsection{Hardware}
As AI increasingly relies on specialized hardware accelerators like GPUs and TPUs to meet computational demands, attackers exploit these components through techniques such as hardware Trojans, side-channel attacks, and hardware-software co-design vulnerabilities. 
 Hardware Trojans implanted in GPUs or TPUs can enable model corruption, backdoor insertion, or information extraction~\citep{xu2021security}.
 Similarly, unencrypted Peripheral Component Interconnect Express (PCIe) traffic has been identified as a novel attack surface, allowing adversaries with physical access to GPUs to extract entire deep neural network (DNN) models using techniques like the Hermes Attack~\citep{zhu2021hermes}.
 Side-channel attacks, such as those targeting GPU memory or PCIe traffic, enable adversaries to extract models or manipulate their behavior~\citep{naghibijouybari2018rendered, wei2020leaky}. 
 These hardware-related vulnerabilities highlight the risks posed by outsourced accelerators and the need for trust in external hardware design.

\section{The AI Development-to-Deployment Lifecycle}
\label{sec:lifecycle}
The lifecycle of an AI system encompasses several distinct stages as shown in Figure~\ref{fig:lifecycle}.
Each phase requires specialized tools, hardware, and software to ensure optimal performance and reliability. 
These stages include data curation, model development, testing, deployment, and ongoing maintenance. 
The threat landscape covers multiple stages in the development to deployed system.
We briefly discuss each here.

\subsection{Data Curation and Preprocessing}
The foundation of any AI system is high-quality data. 
In model development, this stage involves collecting, cleaning, preprocessing, and labeling data to prepare it for training. 
Specialized tools such as data annotation platforms (e.g., Labelbox, Amazon SageMaker Ground Truth) and preprocessing libraries (e.g., Pandas and NumPy) are commonly used.
High-capacity storage systems are often needed to handle very large datasets, which include distributed file systems or cloud storage solutions.
This stage is particularly vulnerable to supply chain attacks due to its reliance on external data sources, third-party tools, and human annotation processes.
Attackers can exploit these dependencies to introduce malicious data, manipulate preprocessing pipelines, or compromise the integrity of feature extraction algorithms. 

Preprocessing transforms raw data into features suitable for training and inference.
From a cyber perspective, attackers can target the preprocessing step to introduce biases, distort features, or manipulate the data pipeline.
Attackers can also compromise preprocessing algorithms, such as normalization, scaling, or feature extraction, to distort the data. 
For example, altering a normalization step or an input sensor could skew numerical features, leading to biased model outputs.
If preprocessing pipelines rely on external libraries or scripts, attackers can inject malicious code to modify the data transformation process.
For instance, a compromised library could add noise to images or truncate text data.
This could be done during development or on the subsequent deployed system.

\subsection{Model Development}
This stage involves designing, training, and validating AI models using curated data.
Developers select algorithms, tune hyperparameters, and optimize performance using specialized frameworks and libraries. 
Popular frameworks such as TensorFlow, PyTorch, and Scikit-learn provide tools for building and training models.
High-performance GPUs (e.g., NVIDIA A100) or TPUs are essential for training large-scale models, especially deep learning systems.

This phase is particularly vulnerable due to the reliance on hardware, external libraries, frameworks, and pre-trained models, as well as the complexity of the development environment.
While these processes enable powerful capabilities, they also introduce unique vulnerabilities outlined in the previous section.

\subsection{Deployment Process}
Once validated, the AI model is integrated into production systems to perform real-world tasks. 
Deployment can occur in cloud environments, edge devices, or in on-premises systems.
Deployment frameworks such as Kubernetes, Docker, and TensorFlow Serving enable scalable and efficient model integration.
Edge devices (e.g., NVIDIA Jetson) are often used for real-time inference, while cloud platforms (e.g., AWS, Google Cloud AI) provide scalable infrastructure for large-scale deployments.
Thus, the deployment phase of an AI model often involves optimization techniques such as quantization and pruning to enable deployment on resource-constrained devices.
While these techniques are essential for efficient deployment, they also introduce unique vulnerabilities that attackers can exploit. 
Transitioning models between formats (e.g., TensorFlow to TFLite to ONNX) during deployment can create security risks due to potential inconsistencies, errors, or malicious modifications.

\subsection{Operation and Maintenance}
The operation and maintenance phase of an AI system is critical for ensuring its long-term reliability, security, and effectiveness. 
During this phase, the deployed model interacts with real-world data, performs inference tasks, and undergoes updates, retraining, and monitoring to adapt to evolving conditions. 
Vulnerabilities arise from model drift, insufficient monitoring, unpatched systems, and insecure update mechanisms.

Over time, the data distribution in the real world may diverge from the data used to train the model, leading to degraded performance. 
Attackers can exploit this drift by introducing adversarial inputs tailored to the new data distribution, further compromising the model’s reliability.
This is an example of a real-time data-poisoning attack that results in reduced accuracy and reliability of the predictions.
This is challenging as the incoming data used for updating a model needs continuous monitoring.

Updates to the underlying infrastructure can be exploited by attackers if the update process is not secure. 
As many AI libraries are consistently updated, many of the same vulnerabilities are present here as they were in the development phase.
Further, AI systems rely on external dependencies such as APIs, cloud services, or third-party libraries. 
Compromised or unreliable dependencies can introduce vulnerabilities during operation.

AI systems often process sensitive data during operation, such as personal information or proprietary business data. 
Attackers may target these systems to exfiltrate data or infer sensitive information through membership inference attacks.

\section{Future of AI Security}
As AI systems become increasingly integrated into critical applications across industries, their security has emerged as a paramount concern. 
The complexity of AI systems, including LLMs, agentic systems, specialized hardware, and intricate software pipelines, introduces a broad attack surface that adversaries can exploit.
From data poisoning and adversarial attacks to hardware Trojans and side-channel vulnerabilities, the threats to AI systems are diverse and evolving. 
The future of AI system security will require a proactive, multi-layered approach that addresses vulnerabilities across the entire lifecycle of AI systems, from development and deployment to operation and maintenance and across the entire AI threat landscape.
Security should be integrated early in the AI lifecycle.

One promising direction is the adoption of red-teaming frameworks for AI systems. 
Red-teaming involves simulating adversarial attacks to identify vulnerabilities before they can be exploited. 
By systematically probing AI systems for flaws, red-teaming enables developers to implement targeted defenses, such as adversarial training, input validation, and robust monitoring. 
Most of the red-teaming frameworks, however, are model-centric.
For example, \citet{perez2022red} demonstrated the effectiveness of red-teaming large language models to uncover weaknesses in their instruction-following mechanisms, such as susceptibility to jailbreaking and adversarial prompts. 
Frameworks like Microsoft’s Counterfit provide tools for automating security testing of AI models beyond LLMs, making it easier to evaluate their resilience against a wide range of attack vectors~\citep{microsoft2023counterfit}.

Frameworks and strategies for assessing the entire AI development and deployment process are beginning to be explored and are addressing the gap between the AI and cybersecurity communities.
The test and evaluation community is instantiating a red-teaming methodology for assessing the safety of AI systems through three assessment pillars~\citep{smith2023test}: (1) robustness and resiliency to adversarial attacks on the AI, (2) cyber assessment of the AI component (hitting on the scope of this paper), and (3) the performance of the AI in potentially contested environments is different from the development environment.
\citet{mauri2022modeling} adapt Microsoft's STRIDE threat modeling framework to identify security risks in AI lifecycle.
STRIDE encapsulates six categories: spoofing, tampering, repudiation, information disclosure, denial of service, and elevation of privilege.
Future efforts along these same lines will help to bridge the gap and improve the security of AI systems.

Another critical area for the future of AI system security is the development of standardized frameworks and protocols for securing hardware-software co-design. 
With the increasing reliance on specialized hardware accelerators, vulnerabilities such as hardware Trojans and unencrypted PCIe traffic pose significant risks.
Collaborative efforts between hardware manufacturers, software developers, and cybersecurity researchers will be necessary to establish trust in the globalized supply chain and ensure the integrity of outsourced components. 
Moreover, the rise of distributed AI systems introduces new challenges, such as covert channel attacks and privacy violations~\citep{xie2024survey}.
Addressing these threats will require innovative solutions, such as differential privacy, secure multi-party computation, and federated red-teaming frameworks. 
By combining proactive security testing, robust design principles, and collaborative research, the future of AI system security can evolve to meet the demands of increasingly sophisticated adversaries while enabling the safe and ethical deployment of AI technologies.

\section{Acknowledgements}
This paper describes objective technical results and analysis. 
Any subjective views or opinions that might be expressed in the paper do not necessarily represent the views of the U.S. Department of Energy or the United States Government.

Sandia National Laboratories is a multimission laboratory managed and operated by National Technology \& Engineering Solutions of Sandia, LLC, a wholly owned subsidiary of Honeywell International Inc., for the U.S. Department of Energy’s National Nuclear Security Administration under contract DE-NA0003525. SAND2025-10822O

\bibliographystyle{apalike}
\bibliography{refs}


\end{document}